\documentclass[twocolumn,prb,superscriptaddress]{revtex4-2}
\usepackage[utf8]{inputenc}
\usepackage{amsmath}
\usepackage{graphicx,amssymb,soul,color,amsmath,bm}
\usepackage{epstopdf,hyperref}
\usepackage[mathcal]{euscript}
\usepackage[normalem]{ulem}
\usepackage{cleveref}
\hypersetup{colorlinks=true,citecolor=blue,linkcolor=magenta}
\sethlcolor{yellow}
\allowdisplaybreaks
\usepackage{xcolor}
\newcommand{\myfont}[1]{{\fontfamily{cmtt}\selectfont{#1}}}

\begin{document}

    \title{Sample generation for the spin-fermion model using neural networks}
    \author{Georgios Stratis}
    \affiliation{Department of Electrical and Computer Engineering, Northeastern University, Boston, Massachusetts 02115, USA}
    \affiliation{Roux Institute, Northeastern University, Portland, Maine 04101, USA}
    
    \author{Phillip Weinberg}
    \affiliation{Department of Physics, Northeastern University, Boston, Massachusetts 02115, USA}
    
    \author{Tales Imbiriba}
    \affiliation{Department of Electrical and Computer Engineering, Northeastern University, Boston, Massachusetts 02115, USA}
    
    \author{Pau Closas}
    \affiliation{Department of Electrical and Computer Engineering, Northeastern University, Boston, Massachusetts 02115, USA}
    
    \author{Adrian E. Feiguin}
    \affiliation{Department of Physics, Northeastern University, Boston, Massachusetts 02115, USA}
    
    \date{\today}
  
    \begin{abstract}
        Quantum Monte-Carlo simulations of hybrid quantum-classical models such as the double exchange Hamiltonian require calculating the density of states of the quantum degrees of freedom at every step. Unfortunately, the computational complexity of exact diagonalization grows $ \mathcal{O} (N^3)$ as a function of the system's size $ N $, making it prohibitively expensive for any realistic system. We consider leveraging data-driven methods, namely neural networks, to replace the exact diagonalization step in order to speed up sample generation. We explore a model that learns the free energy for each spin configuration and a second one that learns the Hamiltonian's eigenvalues. We implement data augmentation by taking advantage of the Hamiltonian's symmetries to artificially enlarge our training set and benchmark the different models by evaluating several thermodynamic quantities. While all models considered here perform exceedingly well in the one-dimensional case, only the neural network that outputs the eigenvalues is able to capture the right behavior in two dimensions. The simplicity of the architecture we use in conjunction with the model agnostic form of the neural networks can enable fast sample generation without the need of a researcher's intervention.
    \end{abstract}

    \maketitle
 
    \section{Introduction}
        \label{sec:intro}
        Strongly correlated materials are characterized by dominant electron-electron interactions \cite{morosan2012,avella2013,avella2015} with electronic and magnetic properties that can potentially be tuned and used for novel technological applications beyond the silicon/semiconductor paradigm. To understand the properties of these systems it is unavoidable to resort to numerical methods such as Quantum Monte Carlo (QMC) \cite{ceperley1986,becca2017}. In models unaffected by the infamous sign problem, QMC can treat very large systems with dozens, if not hundreds, of degrees of freedom. One such scenario is the case of the double-exchange Hamiltonian \cite{zener1951a,anderson1955,degennes1960,kubo1972} (also dubbed, in a different context, as the spin-fermion model) and multi-orbital generalizations. These models have shed light on the physics of manganites with colossal magnetoresistance \cite{dagotto2001, hotta2002,dagotto2003} and high-temperature superconductivity \cite{berg2012}. In such hybrid quantum-classical Hamiltonians, conduction electrons are coupled to classical degrees of freedom, spins in this case. During the QMC sample generation for the spin-fermion model, one needs to calculate the eigenvalues of the electronic Hamiltonian, a task that can be done exactly using matrix diagonalization. Unfortunately, the computational complexity of exact diagonalization scales $ \mathcal{O} (N^3) $ for a matrix of size $ N $, limiting the simulations to a couple of hundred orbitals. Furthermore, the number of updates needed to obtain uncorrelated Monte-Carlo samples can be large close to a phase transition. Thus in order to study larger systems or systems close to a phase transition we need to find faster alternatives to exact diagonalization. There have been various attempts to overcome this issue, such as expanding the density of states using Chebyshev polynomials \cite{furukawa2001,furukawa2004,alvarez2005,weisse2009}, obtaining the Hamiltonian's eigenspectrum corrections after a local update \cite{alvarez2007}, and combining the kernel polynomial method with Langevin dynamics \cite{barros2013}. An alternative approach, which is explored in this paper, is to use data-driven machine learning methods to assist with this computational task.
        
        In the last five years, machine learning has found many applications in quantum physics, such as learning thermodynamics \cite{torlai2016}, finding the ground state of a many-body system \cite{carleo2017}, identifying phase transitions \cite{carrasquilla2017}, conducting quantum state tomography \cite{torlai2018}, and calculating spectral functions \cite{hendry2019, hendry2021}. Moreover, approaches to accelerate Monte-Carlo simulations with machine learning have been explored in the self-learning Monte-Carlo method \cite{liu2017}, using restricted Boltzmann machines \cite{huang2017}, deep \cite{shen2018} and autoregressive neural networks \cite{wu2021}. In this work, we design two neural network models, both featuring simple architectures, that generate samples to determine quantities of interest over a range of temperatures for the spin-fermion/double-exchange model. The first neural network outputs the free energy associated with the quantum degrees of freedom, whereas the second outputs the eigenvalues of the Hamiltonian for a given classical spin configuration. Both networks are trained using a data set that is artificially augmented by taking advantage of the system's translation and rotation symmetries. The manuscript is organized as follows: in \cref{sec:spin_fermion_model} we present the spin-fermion/double-exchange model, in \cref{sec:training} we present the effective models and the training methodology used, in \cref{sec:importance_sampling,sec:neural_net_eff_model} we present our results, and we finally conclude with a discussion in \cref{sec:conclusion}.
                 
	\section{Spin-Fermion Model}
	    \label{sec:spin_fermion_model}
        The spin-fermion model, also known as the double-exchange model, was first proposed by Zener \cite{zener1951a} to explain ferromagnetism in materials with incomplete $ d $-shells and itinerant conduction electrons. It has been extensively studied \cite{anderson1955,degennes1960,kubo1972} due to its relation with colossal magnetoresistance in manganites \cite{dagotto2001,hotta2002,dagotto2003}. In the spin-fermion model, one treats each atom with its $ t_{2g} $ electrons as a classical localized spin $ \mathbf{S}_i \in \mathbb{R}^3 $ at site \textit{i} with unit magnitude $ | \mathbf{S}_i | = 1 $. The $ e_g $ conduction electrons are treated quantum mechanically and are able to move freely interacting only with the classical spin at the site they are located. The Hamiltonian for the spin-fermion model considered in this work is the following single-orbital version:
        \begin{equation}
            \label{eq:spin_fermion_ham}
            \begin{aligned}
                \hat{H}(\mathbf{S}) = & -t\sum_{\langle ij\rangle, \sigma} \left( \hat{c}_{i\sigma}^\dagger \hat{c}_{j \sigma} + \hat{c}_{j \sigma}^\dagger \hat{c}_{i \sigma} \right) \\ &+ \frac{J}{2} \sum_{i,\alpha,\beta, \gamma} S_i^\gamma \hat{c}_{i \alpha}^\dagger \sigma_{\alpha\beta}^\gamma \hat{c}_{i \beta} \,,
            \end{aligned}
        \end{equation}
        where $ \mathbf{S} = \{\mathbf{S}_1, \dots, \mathbf{S}_N\} $ is the classical spin configuration, $t$ is the hopping constant and our unit of energy, $\hat{c}_{i\sigma}^\dagger$ ($\hat{c}_{i\sigma}$) is the fermionic creation (annihilation) operator at the $i$\textsuperscript{th} site for fermion with spin $\sigma \in \{\uparrow, \downarrow\}$, $\langle ij\rangle$ are the pairs of nearest neighbors, $J$ is the interaction strength between the classical spins and the electrons (Hund coupling), $S_i^\gamma$ is the $\gamma$\textsuperscript{th} component of the classical spin at the $i$\textsuperscript{th} site, and $\{\sigma^x, \sigma^y, \sigma^z\}$ are the Pauli matrices.
        
        Due to the fact that our system contains both quantum and classical degrees of freedom, the partition function $ Z $ associated with the spin-fermion model is
        \begin{equation}
        	\label{eq:partition_function}
        	\begin{aligned}
            	Z &= \int d \mathbf{S} \ \text{Tr} \left[ \exp \left( -\beta( \hat{H} - \mu \hat{N})\right) \right] \\
            	  &= \int d \mathbf{S} \prod_\nu \left[1+ e^{-\beta (\epsilon_\nu(\mathbf{S}) - \mu)}\right] \,,
        	\end{aligned}
        \end{equation}
        where $ \beta = \frac{1}{T} $ is the inverse temperature, $\mu$ is the chemical potential, $\hat{N}$ is the number operator, and $\epsilon_\nu (\mathbf{S})$ are the single-particle eigen-energies corresponding to the Hamiltonian with spin configuration $ \mathbf{S} $. The second equality follows from the fact that the electronic degrees of freedom are non-interacting and we are treating them with the grand-canonical ensemble. Given \cref{eq:partition_function}, one can define the probability of finding the system in a given classical spin configuration $\mathbf{S}$ as
        \begin{equation}
        	\label{eq:spin_probability}
        	p( \beta, \mathbf{S} ) = \frac{e^{ -\beta \mathcal{F} ( \beta, \mathbf{S} ) } }{Z} \,,
        \end{equation}
        with $ \mathcal{F} ( \beta, \mathbf{S} ) $ being the fermionic free energy given by
        \begin{equation}
        	\label{eq:free_energy_fermi}
        	\mathcal{F}( \beta, \mathbf{S}) = -\frac{1}{\beta}\sum_\nu\log\left[1+e^{- \beta \left( \epsilon_\nu( \mathbf{S} ) - \mu \right) }\right] \,.
        \end{equation}
        
        Knowing the partition function enables us to determine the expected values for different quantities at various temperatures such as the average energy
        \begin{equation}
        	\label{eq:avg_energy}
        	\begin{aligned}
        		\langle E \rangle = &-\frac{\partial \ln (Z) }{\partial \beta} \\
        		    = &\frac{1}{Z}\int d \mathbf{S} \  e^{-\beta \mathcal{F} ( \beta, \mathbf{S} ) }  \sum_\xi \frac{ \epsilon_\xi (\mathbf{S}) - \mu }{1+ e^{-\beta (\epsilon_\xi ( \mathbf{S}) - \mu)}} \\
        		    = &\frac{1}{Z}\int d \mathbf{S} \  e^{-\beta \mathcal{F} ( \beta, \mathbf{S} ) } \sum_\xi \left[ \epsilon_\xi (\mathbf{S}) - \mu \right] \rho( \beta, \epsilon_\xi, \mu )\,,
    		\end{aligned}
    	\end{equation}
     	and the specific heat
   		\begin{multline}
   			\label{eq:specific_heat}
			C_V = -\frac{ \partial \langle E \rangle }{ \partial T } \\  
			    = k \beta ^2 \Biggl\{ \frac{1}{Z} \int d \mathbf{S} \ e^{-\beta \mathcal{F} ( \beta, \mathbf{S} ) } \biggl[ \biggl( \sum_\xi ( \epsilon_\xi (\mathbf{S}) - \mu ) \rho( \beta, \epsilon_\xi, \mu ) \biggr)^2 \\ 
			 + \sum_\xi \biggl( \epsilon_\xi ( \mathbf{S} ) - \mu \biggr)^2 \rho( \beta, \epsilon_\xi, \mu )( 1 - \rho( \beta, \epsilon_\xi, \mu ) \biggr] - \langle E 	\rangle ^2 \Biggr\} \,,
     	\end{multline}
        where $ \rho( \beta, \epsilon, \mu ) $ is the Fermi-Dirac distribution. Furthermore, using \cref{eq:partition_function,eq:spin_probability} we can calculate the magnitude of the average magnetization
        \begin{equation}
        	\label{eq:avg_magn}
        	| \mathbf{M} | = \frac{1}{Z} \int d \mathbf{S} \ e^{ -\beta \mathcal{F} ( \beta, \mathbf{S} ) } \biggr| \frac{1}{N} \biggl( \sum_i \hat{x} S_i^x + \hat{y} S_i^y + \hat{z} S_i^z \biggr) \biggr| \,,
        \end{equation}
		and the staggered magnetization of our system 
 		\begin{multline}
			\label{eq:avg_stagg_magn}
			| \mathbf{M}_s | = \frac{1}{Z} \int d \mathbf{S} \ e^{ -\beta \mathcal{F} ( \beta, \mathbf{S} ) } \times \\
			 \biggl| \frac{1}{N} \sum_i (-1)^{\sum_{ j=1 }^{d} x_j} \biggl( \hat{x} S_i^x + \hat{y} S_i^y + \hat{z} S_i^z \biggr)  \biggr| \,,
		\end{multline}
		where $ x_j $ in the sum $ \sum_{ j=1 }^{d} x_j $ indicates the site index in the $j$\textsuperscript{th} dimension. The parameters used in this paper favor anti-ferromagnetic order at low temperatures and this behavior should be captured by the staggered magnetization. Based on how we chose to define staggered magnetization, a $| \mathbf{M}_s | = 1$ will indicate a situation where each site has spin with the opposite sign compared with its nearest neighbors.
	
    \section{Model training}
        \label{sec:training}
        In order to compare the different models, in this work we study the spin-fermion model in one-dimension using a system with $ N=20 $ lattice sites, and the two-dimensional counterpart using a system with $ N = 6 \times 6 = 36 $ lattice sites. In both cases, we use interaction strength $ J = -1 $ (ferromagnetic), and chemical potential $ \mu = 0 $. For each temperature we generate training, validation, and testing data sets using three different Metropolis-Hastings Markov chains to eliminate correlation between the data sets. Each data set contains spin configurations with their associated normalized free energy $ \frac{ \mathcal{F( \beta , \mathbf{S} ) } }{N}$ and eigenvalues. For every Monte Carlo update, we use exact diagonalization to determine the free energy of the proposed spin configuration. Each Markov chain consists of a warm-up stage followed by a sample generation stage. During the warm-up stage we generate $ 1000 \cdot N $ samples and determine the acceptance ratio $ r $, which is the number of accepted over the total number of proposed spin configurations. The samples generated during the warm-up stage are not saved. In the sample generation stage, we generate $ \frac{ N_s N } { r } $ samples and save only $N_s$ of them. More specificall, for every $ \frac{ N } { r } $ samples generated we save only the last one. This is done to minimize correlations between samples. The training data set contains $ N_s = 10^4 $ samples, whereas the validation and test data sets contained $ N_s = 10^3 $ samples. We train three different models, an effective Heisenberg model and two neural network models. The effective Heisenberg model is used to benchmark the two neural networks against a well established approach based on physical insight.
        
    \subsection{Effective Heisenberg model}
        \label{sec:linear_model}
        A simple model to approximate the free energy at a given temperature is a linear model inspired by Ruderman-Kittel-Kasuya-Yosida (RKKY) theory \cite{ruderman1954,yosida1957} where each classical spin interacts with every other classical spin via a Heisenberg interaction with a coupling strength that depends on the distance between the two spins \cite{liu2017,kohshiro2021}. The estimator $ \hat{ \mathcal{F} }( \beta, \mathbf{S}) $ of the free energy $  \mathcal{F} ( \beta, \mathbf{S}) $ is given by
        \begin{equation}
            \label{eq:linear_model}
            \hat{ \mathcal{F} }( \beta, \mathbf{S})=  \mathcal{F}_0 + \frac{1}{N}\sum_r J_{r} \sum_i \sum_{j\in \mathcal{N}_{r}} \mathbf{S}_{i}\cdot\mathbf{S}_{j} \,,
        \end{equation}
        where $ \mathcal{F}_0 $ is a constant, $ N $ is the number of sites, $ J_{r} $ is the coupling strength between all classical spins $ j $ in the neighborhood $ \mathcal{N}_{r} $ separated by a distance $ r $ from the classical spin at site $ i $. The constants $ \mathcal{F}_0 $ and $ J_{r} $ are independent of the classical spin configuration, but depend on inverse temperature $ \beta $. This model is linear due to the linear dependence between the free energy and the coupling strength constants $ J_{r} $ and arises from a perturbative treatment of the conduction electrons. We fit the model's parameters using linear regression for a least-squares problem whose solution is given by:
        \begin{equation}
            \label{eq:linear_regression}
            \begin{pmatrix}
                    \mathcal{F}_0 \\
                    \mathbf{J}
            \end{pmatrix}
            = (\mathbf{D}^\intercal \mathbf{D})^{-1}\mathbf{D}^\intercal \mathbf{ F } \,,
        \end{equation}
        where $\mathbf{J}$ is the vector containing the coupling strength constants $ J_r $, $ \mathbf{D}_{m0}\equiv 1 $, $ \mathbf{D}_{mr} \equiv \frac{1}{N}\sum_i \sum_{j\in \mathcal{N}_r{}} \mathbf{S}^{(m)}_{i}\cdot\mathbf{S}^{(m)}_{j} $, and $\mathbf{F}_m$ corresponds to the free energy of training sample $ m $.
        
    \subsection{Neural network models}
        \label{sec:neural_net_models}
		An artificial neural network is a function that maps its input from $ \mathbb{R}^m $ to its output in $ \mathbb{R}^n $. In its simplest form a neural network consists of an input layer, a set of hidden layers, and an output layer where each layer consists of a set of nodes. The $ i^{th} $ node  of the $ j^{th} $ hidden layer, $ h_i^{(j)} $, takes as input the weighted sum of the outputs from the previous layer's nodes $ y_i^{(j)} = \sum_k w^{(j)}_{ik} z^{(j-1)}_k $, and applies a non-linear function $ f(y) $, commonly referred to as the \textit{activation function}, to generate its output $ z_i^{(j)} = f(y_i^{(j)}) $. In this work, without loss of generality, we use the Softplus activation function $ f(y)=\ln(1 + e^y) $. To train a neural network for regression one needs to choose a loss function that compares the true value with the neural network's predicted value. We use the mean absolute error (MAE) $\mathcal{L} = \frac{1}{N} \sum_i |y_i - \hat{y_i}|$ as our loss function and to find its minimum we use the Adam optimizer \cite{kingma2017} with a learning rate that decays over time. At training epoch $ t $ the learning rate is given by $ l_r(t) = l_r(0) \gamma^{-t} $, where $ \gamma $ is the decay constant. We use initial learning rate $ l_r(0) = 10^{-3} $,  decay constant $ \gamma=0.9995 $, and train for a total of $ T = 2 \cdot 10^4 $ epochs. We noticed that in some cases neural networks with extremely low mean squared error generated samples that could accurately describe the system's average energy and specific heat, but the magnitude of the average magnetization departed significantly from the exact results. This issue is addressed by splitting the training set in mini-batches during training, thus introducing some stochasticity in the optimization algorithm. In general, using a large number of mini-batches improves a neural network's performance, but the training time grows as the number of mini-batches increases. As a consequence, we split the training set in as many mini-batches possible while making sure that the training time remains at a reasonable level.
        
        The spin-fermion Hamiltonian has translational and rotational invariance and we take advantage of these two symmetries to augment our training data, with the goal of improving the models' predictive performance. Data augmentation is a well established approach in machine learning due to its implementation simplicity \cite{dao2019}; however, the training time increases since the training set becomes larger. Through data augmentation one tries to teach a neural network the system's symmetries, without imposing explicit constraints on the network's architecture. In this work, for every spin configuration in the training set, we apply a global rotation or translation to get a new spin configuration with the same free energy. Since the rotation invariance is due to a continuous symmetry, in practice we had to choose a discrete set of Euler angles to implement the rotation-based data augmentation. More specifically, we use the twenty-three combinations of Euler angles $ \alpha = \{ 0, \pi \} $, $ \beta = \{ 0, \frac{\pi}{2}, \pi, \frac{3 \pi}{2} \} $, and $ \gamma = \{0, \pi\} $, where the $ \{ \alpha = 0, \beta = 0, \gamma = 0 \} $ combination is excluded since it corresponds to the initial spin configuration. Another approach to utilize a system's symmetries is to construct an equivariant neural network \cite{cohen2016a,wang2020}. An equivariant neural network guarantees that the model obeys the system's symmetries using weight sharing and activation functions that respect those symmetries. An example of an equivariant neural network is a convolutional neural network which respects translation invariance. We opted for data augmentation due to its much simpler and straightforward implementation. Another advantage of data augmentation is the ability to use the same neural network architecture for Hamiltonians that do not share the same symmetries.
        
        At each temperature, we train two different neural networks that share the same architecture, as they are both fully-connected feedforward networks with a single hidden layer, and measure their prediction accuracy using mean squared error on the test data set. The first network, which we refer to as \myfont{N1}, takes as input the classical spin components and outputs the free energy at that specific temperature. The second neural network, which we refer to as \myfont{N1Eigenvalues}, also takes as input the classical spin components, but outputs the Hamiltonian's eigenvalues. For both neural networks, we choose the architecture that combines the smallest mean squared error on a validation set with the least amount of hidden nodes in order to avoid over-fitting and allow for a fast sample generation. For the one-dimensional system, we use $ h = 60 $ hidden nodes and $ n_b = 20$ mini-batches for \myfont{N1} and $ h = 80 $ hidden nodes and $ n_b = 50 $ mini-batches for \myfont{N1Eigenvalues}. For the two-dimensional system, we use $ h=108 $ hidden nodes and $ n_b = 20 $ mini-batches for \myfont{N1} and $ h = 144 $ hidden nodes and $ n_b = 20 $ mini-batches for \myfont{N1Eigenvalues}. In all cases, we observe that the mean squared error on the test data set decreases by starting at high temperatures and moving sequentially to lower temperatures, using the optimized parameters corresponding to the previous temperature's model as the initial parameters for the next temperature's model.
        
\begin{figure}[tbh!]
        	\centering
        	\includegraphics[width=\linewidth]{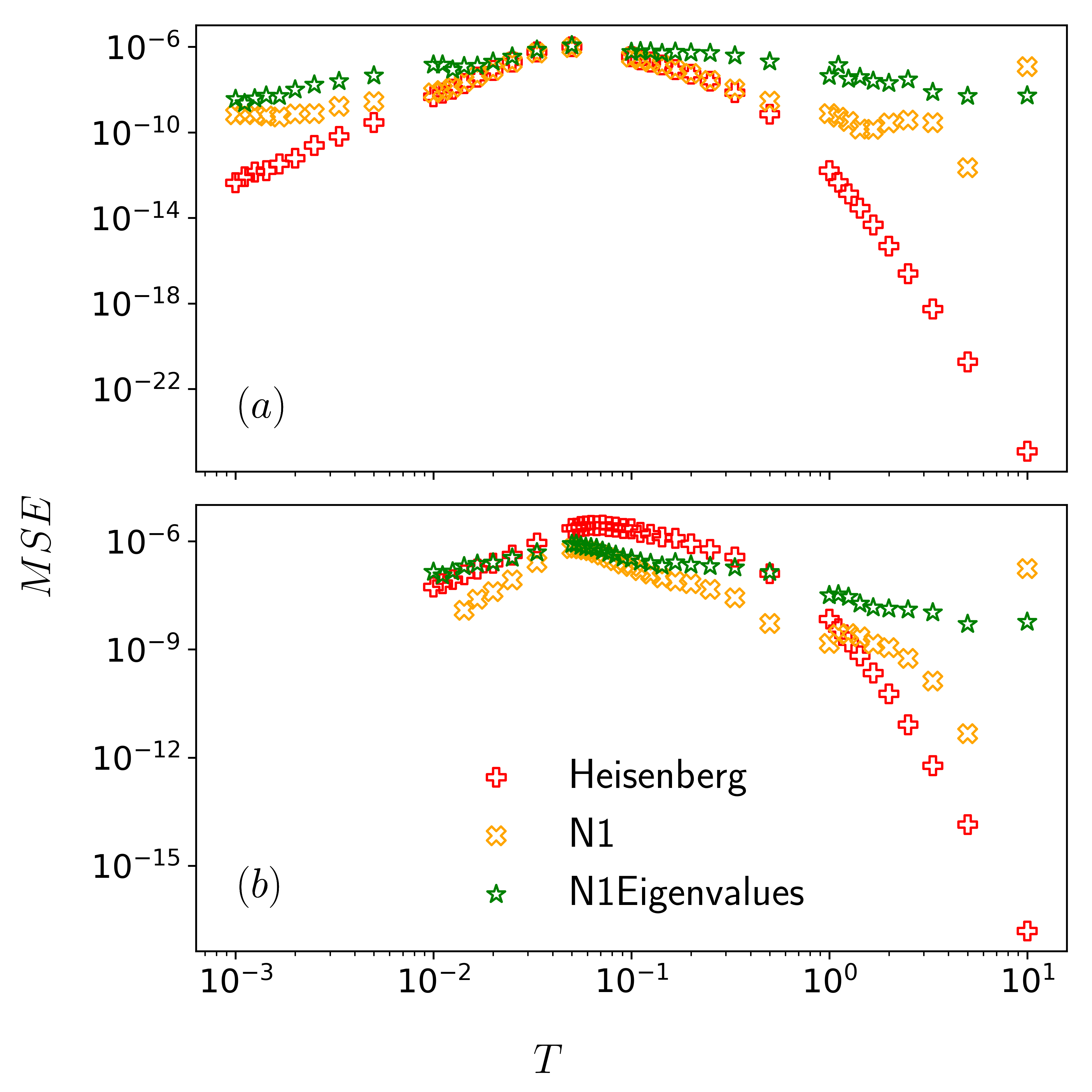}
        	\caption{Mean squared error between the exact  and the predicted fermionic free energy for an one-dimensional lattice $N=20$ sites (panel a) and a two-dimensional lattice $N=6 \times 6= 36$ sites (panel b).}
    	    \label{fig:mse_model_comparison}
        \end{figure}

        For the one-dimensional system with $ N = 20 $ sites, all models have a low mean squared error for the entire range of temperatures, with the linear Heisenberg model having the smallest over the entire temperature range (see \cref{fig:mse_model_comparison}). For the two-dimensional system with $ N = 6 \times 6 = 36 $ sites, all models again have a low mean squared error. The Heisenberg model performs the best at high temperatures whereas the neural networks perform better at lower temperatures (see \cref{fig:mse_model_comparison}). For both systems, the maximum mean squared error reaches $ \sim 10^{-6} $ around  $ \text{T} = 0.05 $ where the average fermionic free energy is $ \mathcal{O}(1) $ indicating that our models are able to predict the fermionic free energy with high accuracy.
        
    \section{Importance sampling} 
        \label{sec:importance_sampling}
        The integrals in \cref{eq:partition_function,eq:avg_energy,eq:specific_heat,eq:avg_magn,eq:avg_stagg_magn} cannot be evaluated in a closed form, requiring us to resort in numerical methods. If the probability distribution from \cref{eq:spin_probability} is known, then the expected value $ \langle O \rangle $ of a function $ O(\mathbf{S}) $ is 
        \begin{equation}
        	\label{eq:expected_value}
	    	\langle O( \beta ) \rangle = \int d \mathbf{S}\ O( \mathbf{S} ) p( \beta, \mathbf{S}) \,,
        \end{equation}
    	and can be approximated with
    	\begin{equation}
    		\label{eq:monte_carlo_expected_value}
    		\begin{aligned}
    			\langle O( \beta ) \rangle \approx \frac{1}{N} \sum_{ i=1 }^{N} O( \mathbf{S}_i )\,,
    		\end{aligned}
    	\end{equation}
    	using spin configurations generated according to \cref{eq:spin_probability}. However, it is computationally expensive to generate samples according to the exact probability distribution governing our system, which is why we resort to using an effective model. More specifically, given the probability distribution according to the effective model $ q( \beta, \mathbf{S})$, the expected value $ \langle O (\beta) \rangle$ is 
        \begin{equation}
        	\label{eq:importance_avg_exact}
   			\langle O (\beta) \rangle = \int d \mathbf{S}\ O( \mathbf{S} )\ q( \mathbf{S} ) \frac{ p( \mathbf{S} ) }{ q( \mathbf{S} ) } \,,
        \end{equation}
		and can be approximated with
		\begin{equation}
			\label{eq:importance_avg}
			\langle O  (\beta) \rangle \approx \frac{ \sum_{ i=1 }^n w( \beta, \mathbf{S}_i) O (\mathbf{S}_i)}{ \sum_{ i=1 }^n w( \beta, \mathbf{S}_i ) } \,,
		\end{equation}
    	where we used
    	\begin{equation}
    		\label{eq:importance_weight}
    		w( \beta, \mathbf{S} ) = e^{ -\beta ( \mathcal{F}( \beta, \mathbf{S} )-\hat{\mathcal{F}}( \beta, \mathbf{S} ) )} \,,
    	\end{equation}
    	with $ \mathcal{F}( \beta, \mathbf{S} ) $ and $ \hat{\mathcal{F}}( \beta, \mathbf{S} ) $ being the fermionic free energy due to the exact and effective models respectively. This process is referred to as \textit{importance sampling} in literature \cite{robert2004}. It is worhtwhile noting that the estimator in \cref{eq:importance_avg} is asymptotically unbiased \cite{bugallo2017}. Calculating \cref{eq:importance_weight} requires us to diagonalize the Hamiltonian for a given spin configuration. One might reasonably ask what computational gains do we get by using an effective model if we still need to perform exact diagonalization for importance sampling? The computational savings come during the de-correlation stage of the Metropolis-Hastings algorithm where for every $ \frac{N}{r} $ spin configurations generated we record only the last one so using an effective model for all those unrecorded update steps reduces the time needed for sample generation.
	
\begin{figure}[tbh!]
			\centering
			\includegraphics[width=\linewidth]{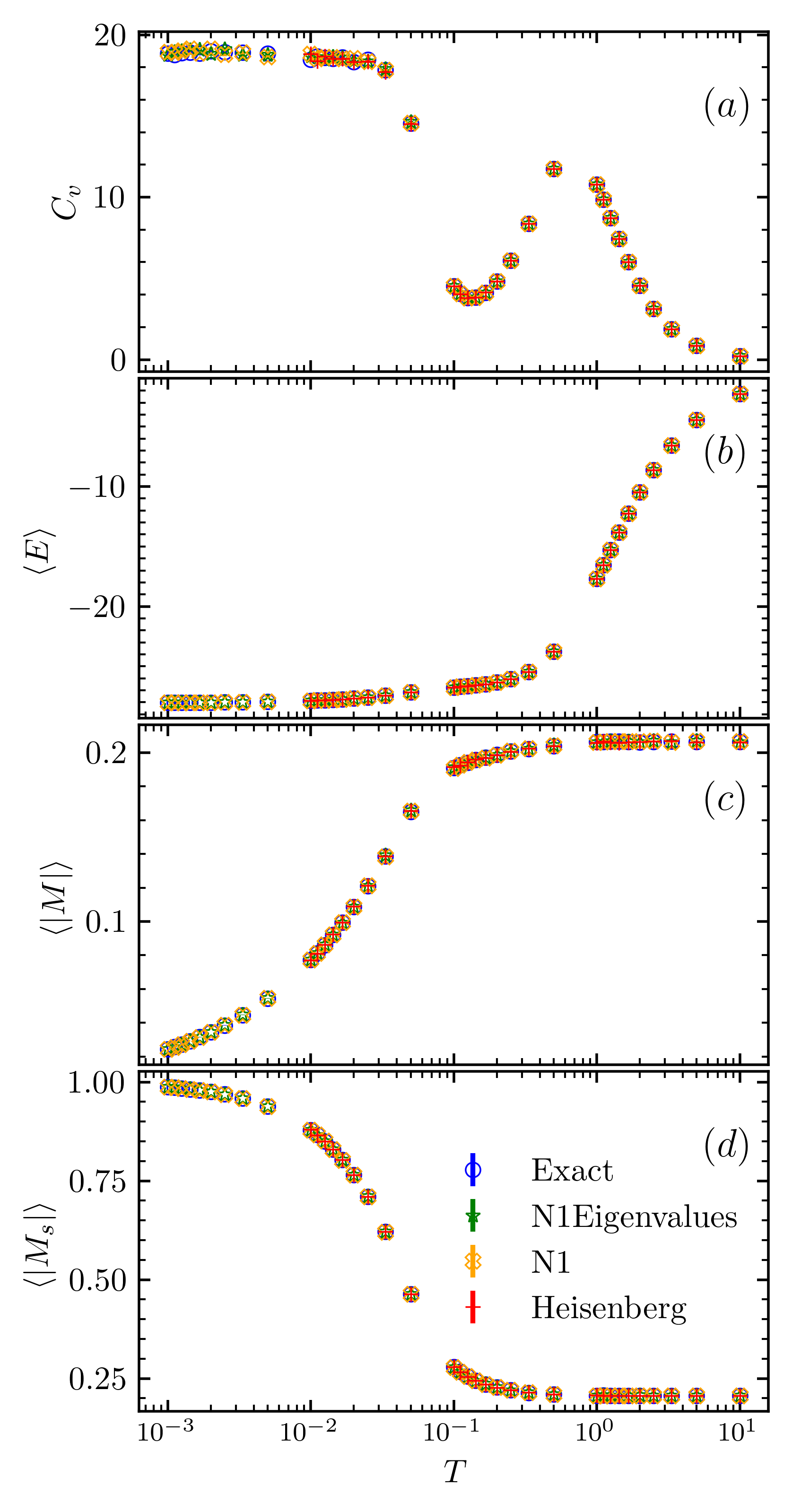}
			\caption{Specific heat (panel a), average energy (panel b), magnitude of average magnetization (panel c), and staggered magnetization (panel d) for an one-dimensional system with $ N=20 $ lattice sites. We generated $10^5$ samples using exact diagonalization (blue circles), \myfont{N1Eigenvalues} (green stars), \myfont{N1} (orange crosses), and Heisenberg model (red pluses). All models are in excellent agreement with the exact results.}
			\label{fig:1d_results}
		\end{figure}

        For the one-dimensional system ($ N=20 $ lattice sites), the effective models generate samples that accurately describe the average energy, specific heat, the magnitude of average magnetization, and the staggered magnetization as can be seen in \cref{fig:1d_results}. For the Heisenberg model, we do not generate samples below $ T = 10^{-2} $ because the sample generation script requires approximately the same amount of time to generate samples as exact diagonalization, a fact we attribute to our code not being properly optimized.
		
\begin{figure}[tbh!]
   			\centering
   			\includegraphics[width=\linewidth]{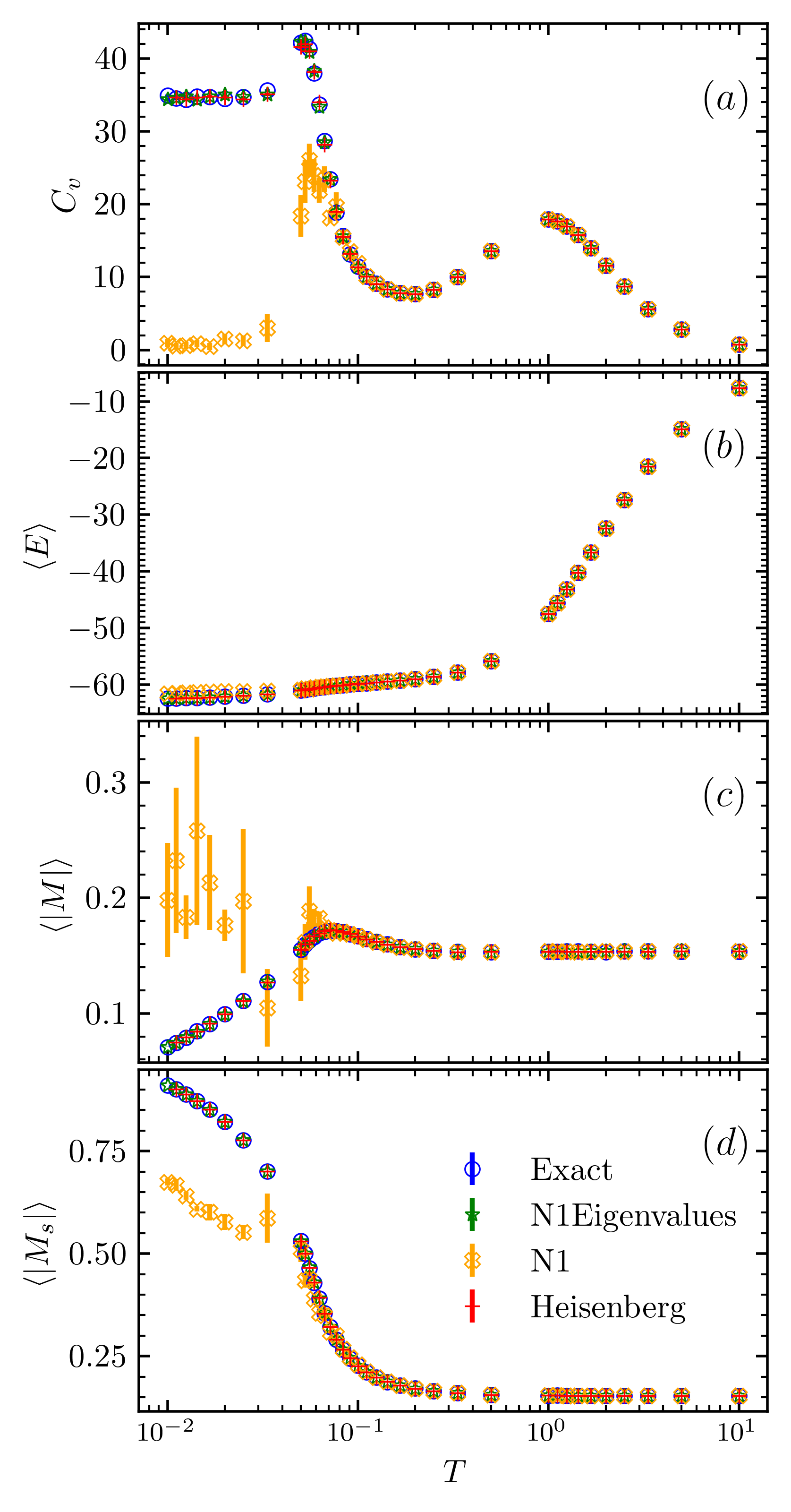}
            \caption{Specific heat (panel a), average energy (panel b), magnitude of average magnetization (panel c), and staggered magnetization (panel d) for a two-dimensional system with $ N = 6 \times 6=36$ lattice sites. We generated $10^5$ samples using exact diagonalization (blue circles), \myfont{N1Eigenvalues} (green stars), \myfont{N1} (orange crosses), and Heisenberg model (red pluses). The \myfont{N1} model fails to properly describe the system's behavior in the low temperature regime.}
            \label{fig:2d_results}
        \end{figure}

		 The situation is different for the two-dimensional case ($ N = 6 \times 6 = 36 $ lattice sites). All models generate samples that are able to accurately describe the average energy, but the \myfont{N1} model fails to capture the correct behavior for the other three quantities in the region of $ T \lesssim 0.1 $ as can be seen in \cref{fig:2d_results}. This seems counterintuitive given the extremely low mean squared error that \myfont{N1} has as can be seen in \cref{fig:mse_model_comparison}, but this is an indication of overfitting. We have tried increasing both the number of mini-batches used during training and the number of hidden nodes for the \myfont{N1} model; however the results remained practically the same. 

    \section{Neural networks as effective models}
        \label{sec:neural_net_eff_model}
    	Lastly, we examine the possibility of completely eliminating exact diagonalization from the calculation of the expected values. That is equivalent to using \cref{eq:monte_carlo_expected_value} where the spin configurations are drawn from the probability distribution derived from the effective models. In other words, neural networks can be seen as {\it non-perturbative} effective models obtained by tracing out over the electronic degrees of freedom.
    	
\begin{figure}[tbh!]
    		\centering
    		\includegraphics[width=\linewidth]{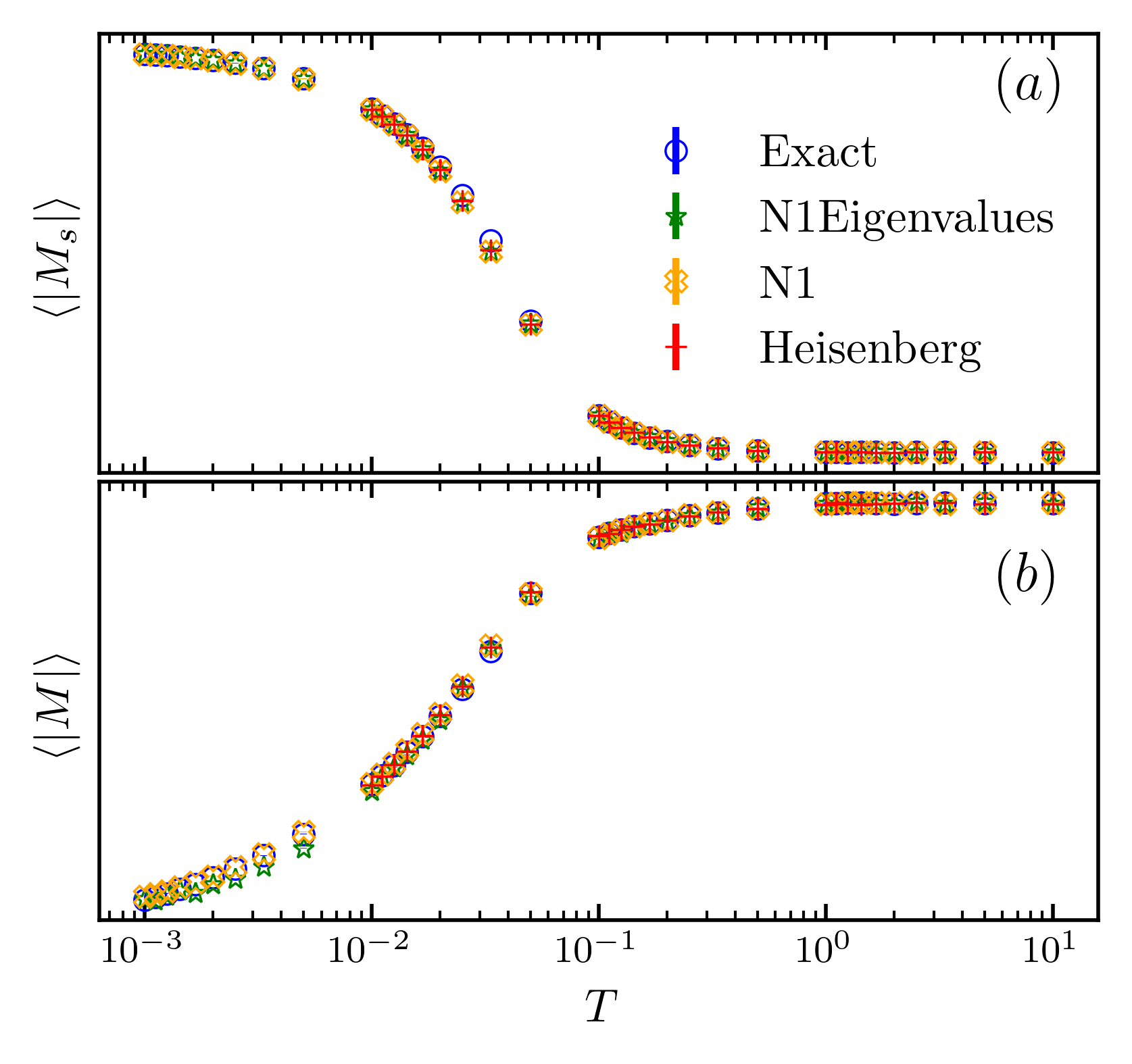}
    		\caption{Staggered magnetization (panel a) and magnitude of average magnetization (panel b) for the one-dimensional system ($ N = 20$ lattice sites) without implementing importance sampling. All three models maintain the same performance as shown in \cref{fig:1d_results}.}
    		\label{fig:1d_noimportancesampling}
    	\end{figure}

\begin{figure}[tbh!]
    		\centering
    		\includegraphics[width=\linewidth]{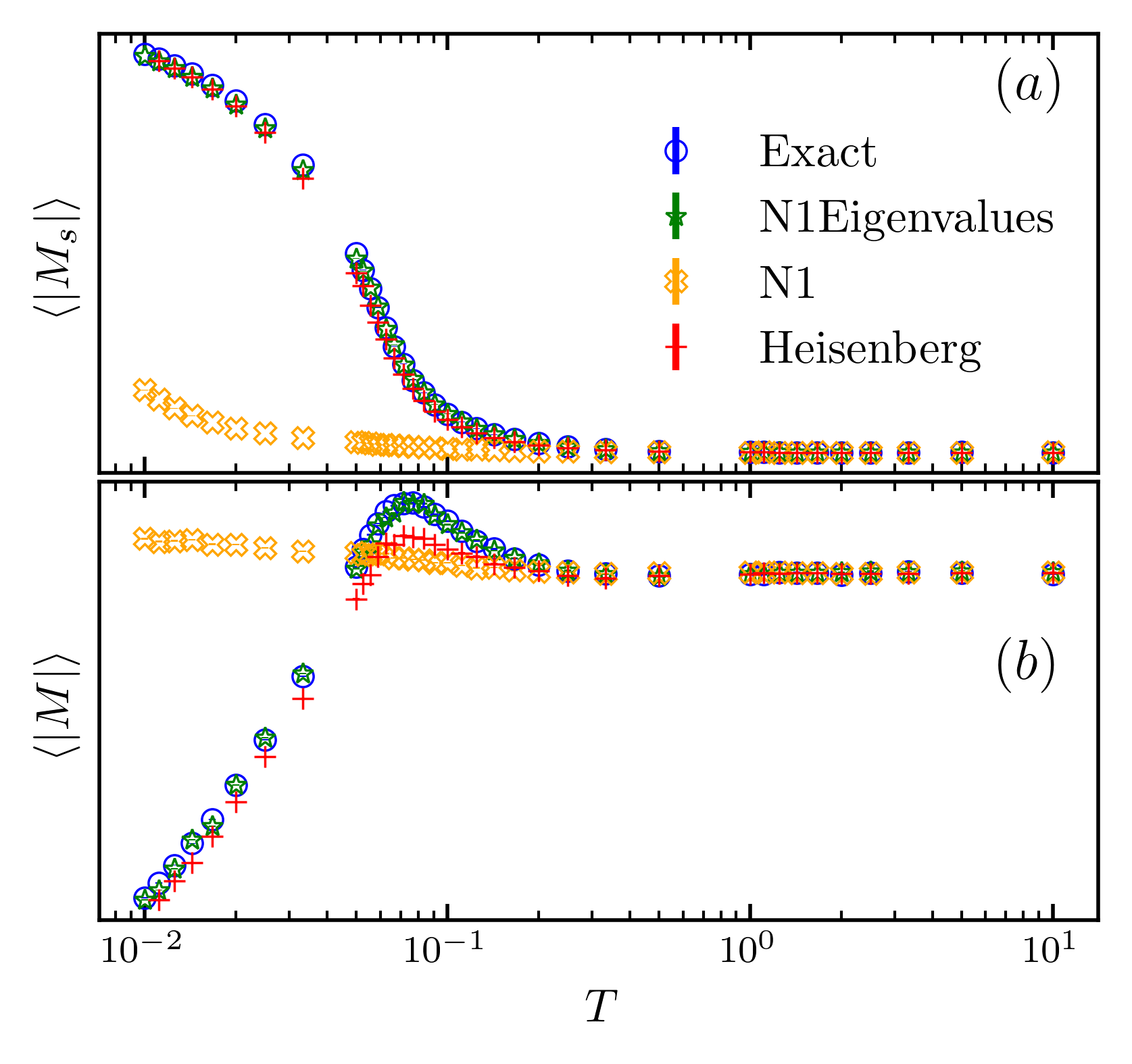}
    		\caption{Staggered magnetization (panel a) and magnitude of average magnetization (panel b) for the two-dimensional system ($ N = 6 \times 6 = 36 $ lattice sites) without implementing importance sampling. The \myfont{N1Eigenvalues} maintains the same performance as shown in \cref{fig:2d_results}, whereas the Heisenberg model shows a slight disagreement in the region around $ T \sim 0.1 $. The \myfont{N1} model completely fails for temperatures below $T \sim 0.2$.}
    		\label{fig:2d_noimportancesampling}
    	\end{figure}

    	To test this idea, we calculate the observables related to the magnetization state of the system since both the average energy and specific heat require knowledge of the eigenspectrum which cannot be accessed with the \myfont{N1} and Heisenberg models. Having said that, we could still use \myfont{N1Eigenvalues} for determining these two quantities since its output is the eigenspectrum of the Hamiltonian. For the one-dimensional system (see \cref{fig:1d_noimportancesampling}), all three models remain in agreement with exact diagonalization maintaining the same performance as in \cref{fig:1d_results}. For the two-dimensional system (see \cref{fig:2d_noimportancesampling}), only \myfont{N1Eigenvalues} remains in agreement with exact diagonalization across all temperatures. Interestingly, the Heisenberg model remains in agreement with exact diagonalization for most temperatures, and struggles only in the region around $ T \sim 0.1 $. The \myfont{N1} completely fails to capture the correct behavior for temperatures below $ T \sim 0.2 $, which is precisely the regime where the physics is dominated by the energy spacing. On the other hand, it is reasonable to expect that \myfont{N1Eigenvalues} will be better to retain its high fidelity even in the absence of importance sampling since the network approximates the mapping of each spin configuration to its corresponding eigenspectrum, a task which might preserve a significant amount of information relevant for describing the system. In contrast, both the Heisenberg and \myfont{N1} models try to approximate the fermionic free energy, a scalar quantity obtained by integrating over the spectrum, a process that might cause significant information loss.
    	    	
    \section{Conclusion}
        \label{sec:conclusion}
        Neural networks are ideal surrogates to replace computationally intensive steps in Quantum Monte Carlo, given their remarkable flexibility to approximate arbitrary functions of interest. This advantage is counterbalanced by the numerical cost required to train the models. One solution is to exploit a system's symmetries and augment the existing training data set, thus creating a bigger data set without the overhead of the exact method. An added benefit of data augmentation is that we are actively encouraging the neural network to learn the system's symmetries that the model should also obey. In this work, we train two neural network models using data augmentation and demonstrate their ability to predict the free energy for the spin-fermion model in one and two dimensions. The two neural network models differ fundamentally in two aspects: (i) the quantity they learn, and (ii) how they are used as effective models to generate samples. The \myfont{N1} model learns the free energy and is used in combination with importance sampling which requires to diagonalize the Hamiltonian at each measurement step. The \myfont{N1Eigenvalues} model learns the energy spectrum (or density of states) of the Hamiltonian, and can reproduce the exact results with high level of accuracy without using importance sampling, making the simulations considerably more efficient and faster. We compare the two neural networks against an effective Heisenberg model and find that in one dimension all three models have comparable performance, whereas in two dimensions only the \myfont{N1Eigenvalues} and the Heisenberg model describe correctly the system's behavior across all temperatures. Unlike the Heisenberg model, the \myfont{N1Eigenvalues} model is not constrained to have a specific functional form and is able to find an appropriate approximation to the energy spectrum. Lastly, since the energy spectrum is temperature independent, a neural network that learns the energy spectrum, such as the \myfont{N1Eigenvalues} model, could be trained at a high temperature where it is cheap to generate a large training data set and then used for all temperatures. Such an approach will reduce the cost associated with training allowing the study of larger systems at low temperatures. 
               
    \section{Acknowledgments} 
        \label{sec:acknowledgments}
        We acknowledge the support we received from NU TIER1 FY21 and the National Science Foundation under Awards ECCS-184583 (TI and PC) and DMR-2120501 (AEF). GS was supported by the Roux Institute at Northeastern University and the Harold Alfond foundation.
        
    \bibliography{spin-fermion}
    
\end{document}